\documentclass[aps,nofootinbib]{revtex4-1}

\usepackage{bm}
\usepackage{hhline}

\usepackage{graphicx}
\usepackage{amsmath, amsthm, amssymb,amsfonts}

\usepackage{color}

\begin{document}

\preprint{APS/123-QED}

\title{``Ideal'' tearing and the transition to fast reconnection in the weakly collisional MHD and EMHD regimes}

\author{{Daniele Del Sarto}}
 \affiliation{  Institut Jean
Lamour, UMR 7198 CNRS - Universit\'e de Lorraine, BP 239 F-54506 Vandoeuvre-le-Nancy, France}
 \email{daniele.del-sarto@universite-lorraine.fr}
\author{Fulvia Pucci}
\affiliation{ Dipartimento di Fisica, Universit\`a di Roma-Tor Vergata, Roma, Italy}%
\author{Anna Tenerani}%
\affiliation{Earth, Planetary and Space Sciences, UCLA, Los Angeles, USA}
\author{Marco Velli}%
\affiliation{Earth, Planetary and Space Sciences, UCLA, Los Angeles, USA}

%
%


\date{\today}

\begin{abstract}
This paper discusses the transition to fast growth of the tearing instability in thin current sheets in the collisionless limit where electron inertia drives the reconnection process. It has been previously suggested that in resistive MHD there is a natural maximum aspect ratio (ratio of sheet length and breadth to thickness) which may be reached for current sheets with a macroscopic length $L$, the limit being provided by the fact that the tearing mode growth time becomes of the same order as the Alfv\'en time calculated on the macroscopic scale  (\emph{Pucci and Velli (2014)} \cite{id_tearing}). For current sheets with a smaller aspect ratio than critical the normalized growth rate tends to zero with increasing Lundquist number $S$, while for current sheets with an aspect ratio greater than critical the growth rate diverges with $S$. Here we carry out a similar analysis but with electron inertia as the term violating magnetic flux conservation: previously found scalings of critical current sheet aspect ratios with the Lundquist number are generalized to include the dependence on the ratio $d_e^2/L^2$ where $d_e$ is the electron skin depth, and it is shown that there are limiting scalings which, as in the resistive case, result in reconnecting modes growing on ideal time scales. Finite Larmor Radius effects are then included and the rescaling argument at the basis of ``ideal'' reconnection is proposed to explain secondary fast reconnection regimes naturally appearing in numerical simulations of current sheet evolution.
\end{abstract}

\maketitle

\section{Introduction}\label{Intro}
Magnetic reconnection is thought to be the mechanism underlying many explosive phenomena observed in both space and laboratory plasmas, ranging from magnetospheric substorms, to solar flares and coronal mass ejections, to the sawtooth crashes observed in tokamaks. The classic picture of reconnection involves current sheets, most often assumed to be planar-like and concentrated more narrowly in the third dimension. Often, a guide magnetic field lies within the current sheet itself, so that the actual three-dimensional field does not vanish in the sheet. Different models for reconnection occurring in such quasi-2D configurations have been developed, two prominent, different examples being the  Sweet-Parker (SP) stationary reconnection scenario and the spontaneous reconnecting modes naturally developing due to the tearing instability of the current sheet itself. \emph{Biskamp (1986)} \citep{Biskamp1986} first pointed out the important role played by the current sheet aspect-ratio in determining whether a stationary reconnection configuration could be reached. He found, via numerical simulations, that the SP current sheet could become unstable to reconnecting modes once a critical value of the Lundquist number (estimated on the current sheet length or breadth, $L$) of about $S\simeq 10^4$ was exceeded. A detailed examination of the stability of the SP configuration led to the definition of the plasmoid-chain instability  \citep{Loureiro_plasmoid},  reminiscent of the plasmoid-induced reconnection concept and fractal reconnection models introduced by  \emph{Shibata et al. (2001)} \citep{Shibata}. Recently, \emph{Pucci and Velli (2014)} \citep{id_tearing} have pointed out that the divergence of the growth rate of the plasmoid chain instability in the limit of large Lundquist number within resistive MHD implies that current sheets should never elongate sufficiently to achieve the SP aspect ratio. They have shown that a critical aspect ratio separates slowly unstable current sheets (with growth rate scaling as a negative, fractional exponent of the Lundquist number) from violently unstable ones (growth rates scaling with a postive power of $S$). They dubbed the instability of the critically unstable current sheet ``ideal tearing" (hereafter IT), because the growth rate, normalized to the Alfv\'en time {\it along the sheet} $L$, becomes of order unity, and independent of the Lundquist number itself.

The large predicted growth rates and the presence of critical values for dimensionless numbers such as current-sheet aspect ratio make the described instabilities good candidates to understand and model the mechanisms behind observed fast reconnection phenomena \citep{Velli2}. Indeed, to date there is no agreed theoretical explanation for the fast time scales over which reconnection events develop in nature, nor for their triggering, while evidence from both experiments and numerical simulations points to the importance of small scale formation and kinetic effects, \citep{Daughton,Moser,Biancalani} which are theoretically expected to lead to Alfv\'enic (or ``ideal") reconnection in 3D configurations  as well \citep{Boozer_space}. Moreover, numerical simulations of tearing mode instabilities have identified a secondary, nonlinear, increase of the reconnection rate, that has been sometimes interpreted in terms of a nascent plasmoid-unstable SP regime \citep{Loureiro_2005,Ali} or generically a secondary ``explosive reconnection'' regime \citep{Biancalani}. A nonlinear increase of the reconnection rate on ideal, Alfv\'enic time-scales was also numerically measured by \emph{Yu. et al. (2014)}  \citep{Yu} in simulations of low mode-number reconnection instabilities. Given the recent developments of the theory of large-aspect ratio current sheet instabilities, it is important to understand whether such augmented fast reconnection rates may indeed be interpreted as fast secondary instabilities of the nonlinearly generated current sheets stemming from the primary reconnection event.  Specifically, given that kinetic and two-fluid effects easily become dominant compared to classical, collisional resistivity at small spatial scales, it seems timely to see whether and how such effects modify the transition to an IT regime. 

The present paper focuses on the extension of the IT scaling arguments to weakly collisional regimes where reconnection is mediated by electron inertia effects, and on whether such generalized IT regimes might explain the nonlinear occurrence of fast exponentially growing reconnection rates. We will consider both the incompressible reduced MHD (RMHD $-$ see e.g. \citep{Zank}) and electron MHD (EMHD \citep{Kingsep}) frequency ranges, where the perturbations are dominated by Alfv\'en and whistler modes respectively. The formal similarity between RMHD and EMHD reconnection in slab geometry, previously discussed in \citep{KH,MPLB}, allows a unified treatment for the onset of IT in an electron-inertia driven framework.

Electron inertia has long been considered the most promising alternative to standard resistive reconnection thanks to its greater weight with respect to resistivity in the generalized Ohm's law of  quasi-collisionless plasmas \citep{Coppi1,Coppi2,Wesson,Porcelli}.  Astrophysical and thermonuclear fusion plasmas are examples of such systems, since their particle mean free path tipically exceeds  the characteristic hydrodynamic lengths by order(s) of magnitude. In general, interspecies collisions  may be neglected with respect to inertial terms when the characteristic ion-electron collision frequency is negligible with respect to the inverse time scale of the phenomena considered \citep{Ottaviani1,Porcelli1,Hosseinpour_1}. The inertial slab RMHD regime we focus on here has indeed been widely used to model basic features of magnetic reconnection in tokamak devices, for which the strong guide field approximation, of which we consider the 2D-geometry limit, was first devised, as well as for astrophysical applications.  In EMHD the neglect of collisional resistivity is even more justified,  which is why EMHD reconnection is mostly studied in purely inertia-driven regimes (see \citep{Hosseinpour_1} for a discussion of the transition from resistive to inertial EMHD). Because of the large characteristic frequencies involved, EMHD provides a natural framework for collisionless reconnection. The relation between the convection electron flow and the magnetic field, typical of the  EMHD regime,  plays a prominent role in explaining the quadrupolar structure of the out-of-plane magnetic field  \citep{Bian}, which is often recognized as a distinctive signature for the \emph{in situ} detection of magnetospheric reconnection \citep{Oieroset}. \emph{Rogers et al. (2001)} \cite{Rogers} also adopted the incompressible, inertia-less, collisionless EMHD model to explain the opening-up of the reconnection layer in 2D simulations with no guide field. We finally note that the present paper does not cover the framework of the so-called Hall- or whistler- mediated reconnection  (Appendix \ref{Ohm_Hall}), especially relevant to the magnetopause environment \citep{GEM,Vaivads}, and which is known to provide prominent examples of fast reconnection rates weakly dependent from both resistivity \citep{Mandt} and electron inertia \citep{Biskamp1}. This will be considered in future works.

The paper is structured as follows. In Sec.\ref{Ideal_tearing} we summarize the re-scaling arguments leading to the concept of ``ideal tearing". In Sec.\ref{Model} we introduce  the model equations for reconnection in the RMHD and EMHD regimes and the relevant dispersion relations (Sec.\ref{linear}). In Sec.\ref{Results} we extend the IT paradigm first to the inertial RMHD and EMHD reconnection regimes (Sec.\ref{Results_collisionless}) and then to include finite Larmor radius (FLR) effects (Sec.\ref{Results_FLR}). We then discuss these results (Sec.\ref{Discussion}) by comparing the role of inertia to that of resistivity in different natural and laboratory plasmas (Sec.\ref{inertia}), and by considering an application of the IT model to collisionless steady reconnecting current sheets  (Sec.\ref{steady}). Then, in Sec.\ref{secondary}  we discuss how the re-scaling argument might explain explosive reconnection regimes nonlinearly observed in simulations of magnetic reconnection. Sec.\ref{summary} provides a summary and conclusion and in the Appendix \ref{App:model_equations} we recall the derivation of the model equations both from a two-fluid model and compared with the generalized Ohm's law (Sec.\ref{Ohm_Hall}).

\section{The ideal tearing model}\label{Ideal_tearing}
Consider a current sheet of length $L$ and thickness $a$.  As MHD is scale-free, in the classical tearing mode theory it is customary to take  the width $a$ as normalization length, since typically $L/a>1$ and $a$ is the only characteristic length defined by the (usually 1D) equilibrium profile. However, when dealing with thin sheets with $a$ arbitrarily small, the distinction between $L$ and $a$ becomes important, as the tearing mode growth rate is only small when measured with respect to the ``ideal" Alfv\'en timescale based on $a$, but can become large when measured with respect to a macroscopic scale $L >>a$ (the basic idea behind the plasmoid instability and IT, detailed below).  From now on, we will label quantities normalized to the scale $L$ with the apex ``$*$'', using  standard notation for non-dimensional quantities defined in terms of the (possibly microscopic) shear-scale $a$. 

In this notation, the classical linear reconnecting mode on Harris-type current sheets has a maximal growth rate scaling as $\gamma_M \tau_{_A} \sim S^{-1/2}$ where the Lundquist number $S = aV_{_A}/\eta_m$ and $\tau_{_A} = a/V_{_A}$, with $V_{_A}$ the Alfv\'en speed  based on the characteristic magnetic field strength far from the sheet. In the SP case, predicated on the renormalized Lundquist number $S^* = LV_{_A}/\eta_m$, one finds immediately that $\gamma_M \tau_{_A}^* = \gamma_M L/V_{_A} \sim {S^*}^{1/4}$, i.e. a growth rate which diverges with the macroscopic Lundquist number $S^*$. \emph{Pucci and Velli (2014)} \citep{id_tearing}, aiming to resolve this paradox, incompatible with the ideal MHD limit, studied large-aspect ratio current sheets with $L/a$ scaling as a positive fractional power of the Lundquist number $S^* = LV_{_A}/\eta_m \gg 1$. They showed that when  a threshold $L/a\sim (S^*)^{\alpha}$ ($1/2>\alpha > 0$) is reached, the resistive tearing mode growth rate $\gamma_M \tau_{_A}^*$ becomes of order unity and independent of $S^*$.  This regime was named ``ideal tearing'', in contrast to the CT theory in which 
the growth rates scale as a negative power of the $a$ -normalized Lundquist number $S$. The large aspect ratio limit allowed \cite{id_tearing} to evaluate the characteristic CT reconnection rate through the fastest  growing mode, from which the value  
$\alpha=1/3$ was obtained, leading to the conclusion that SP current sheets should not form at large $S^*$ (different equilibrium profiles may induce small deviations from this value \citep{equilibri}). The renormalization in fact gives 
\begin{equation}\label{eq:id_tearing_Harris}
\gamma_M \tau_{_A}^*\sim (S^*)^{-1/2}(L/a)^{3/2}\end{equation}
and the clock whose rate defines the reconnection speed enters this renormalized theory through $\tau_{_A}^*$ which depends itself on $L$, i.e., the clock set on the ideal scale $L$ results slower by a factor $a/L$ (or, as we shall see, $(a/L)^2$ in the EMHD regime) than the clock  with which the reconnection rate is measured in the CT theory: it is thus always possible to find a critical exponent $\alpha>0$ such that $\gamma_M\tau_{_A}^*\simeq 1$ once the condition  $(L/a)\sim (S^*)^{\alpha}$ is imposed. In other words, the tearing-mode theory, under the assumption of a current sheet whose aspect ratio scales as a power of the (small) non-ideal parameter $\varepsilon^*$ which allows  reconnection,  say $a/L\sim (\varepsilon^*)^\alpha$, can explain the transition to fast reconnection if the value of $\alpha$ is such that the growth rate of the instability is independent from $\varepsilon^*$ itself. Notice however that the IT criterion may be applied in principle to any reconnection unstable aspect ratio $L/a$,  if $L$ is large enough with respect to $a$. It is e.g. the case of tearing unstable current sheets, nonlinearly developed by primary reconnection events, which we will consider later.  We now consider how this happens once electron inertia first, and FLR-type effects second, are taken into account. 

\section{Model equations}\label{Model} 
We restrict our analysis to a 2D system in the $(x, y)$ plane, and assume for simplicity an electron-proton plasma. 
Consider the incompressible equations in slab-geometry. 
We adopt the standard  ``poisson-bracket'' representation $[f,g]\equiv\partial_x f\partial_y g-\partial_y f\partial_x g={\bf e}_z\cdot({\bm \nabla}f\times{\bm \nabla}g)$. The velocity stream functions $\varphi$ and $b$ are such that 
${\bm U}_\perp=-{\bm\nabla}\varphi\times{\bf e}_z$  in RMHD and  ${\bm u}^e_\perp=-{\bm\nabla}b\times{\bf e}_z$  in EMHD (see below),   where ``$\perp$'' stands for components in the $(x, y)$ plane, and ${\bm u}^e$ and ${\bm U}$ are the electron and bulk plasma velocities, respectively. Analogously, the magnetic stream function $\psi$ is defined through ${\bm B}={\bm \nabla}\psi(x,y)\times{\bf e}_z+(B_0+b(x,y)){\bf e}_z$, with $B_0$ uniform in space. We assume an equilibrium in-plane magnetic field ${\bm B}_\perp^0=B_y^0(x/a){\bf e}_y$ with $B_y^0(x/a)= \partial_x \psi_0(x/a)$. Equilibrium quantities are labeled with ``$0$'', and we introduce the fields  $F\equiv \psi-d_e^2\nabla^2\psi$ and $W\equiv b-d_e^2\nabla^2b$. Here $d_e= c/\omega_{pe}$ is the electron-skin-depth. 
		
Using $a$ as the reference length and characteristic quantities $B_\perp^0$ and $n_0$ for magnetic field and densities, the model equations may then be written in non-dimensional form  either as:
\begin{equation}\label{eq:OhmRMHD}
\frac{\partial}{\partial t}F+[\varphi,\,F]=\rho_s^2[\nabla^2\varphi,\psi]+
S^{-1}{\nabla}^2\psi
\end{equation}
\begin{equation}\label{eq:RMHD_U}
\frac{\partial}{\partial t}\nabla^2\varphi+[\varphi,\,\nabla^2\varphi]=[\psi,\,\nabla^2\psi] +R^{-1}{\nabla}^4\varphi\, ,
\end{equation}\\
valid in the RMHD frequency range, or
\begin{equation}\label{eq:OhmEMHD}
\frac{\partial}{\partial t}F+[b,\,F]=S_{_{Emhd}}^{-1}{\nabla}^2\psi\end{equation}
\begin{equation}\label{eq:EMHD_U}
\frac{\partial}{\partial t}W+[b,\,W]=[\psi,\,\nabla^2\psi] +S_{_{Emhd}}^{-1}{\nabla}^2b\, ,
\end{equation}
valid in the EMHD frequency range. 

In the above, time is normalized to  $\tau_{{_A}}\equiv (a/ d_i)\Omega_{i}^{-1}$ in RMHD, where $\Omega_{i}$ is the ion cyclotron frequency and $d_i\equiv \sqrt{m_i} c/(\sqrt{m_e}\omega_{pe})$ is the ion-skin depth ($\omega_{pe}$ being the usual plasma frequency and with obvious notation for the masses); 
in EMHD time is normalized to  the inverse of the whistler frequency, $\tau_{{_W}}\equiv (a/d_e)^2\Omega_{e}^{-1}=(a/d_i)^2\Omega_{i}^{-1}$. 
The other parameters on which the tearing reconnection rate depends are the ion-sound Larmor radius, also non-dimensionalized with $a$ i.e. $\rho_s\equiv c_{is}/a\Omega_i$, where $c_{is}$ is the ion sound speed, i.e. the thermal speed based on electron temperature and ion mass;  $R\equiv (\nu_{ii}\tau_{_{A}})^{-1}$ (Reynold's number) with $\nu_{ii}$ the ion-ion viscosity; $S\equiv \tau_{_D}/\tau_{_{A}}$ (Alfv\'enic Lundquist number) and  $S_{_{Emhd}}\equiv \tau_{_D}/\tau_{_{_W}}$ (EMHD Lundquist number) with $\tau_{_D} = 4\pi a^2/(\eta c^2)$ the resistive diffusion time ($\eta$ is the scalar resistivity). The physical meaning of the terms of Eqs.(\ref{eq:OhmRMHD}-\ref{eq:EMHD_U}) and their relation to both the two-fluid model equations and the generalized Ohm's law are discussed in Appendix \ref{App:model_equations}. 

Note that, calling $L_{_{MHD}}$ and $L_{_{EMHD}}$ the normalization lengths in RMHD and EMHD,  
the inequality
\begin{equation}
\frac{\tau_{_{W}}^*}{\tau_{_A}^*}= \left(\frac{L_{_{EMHD}}}{d_i}\right)
\left(\frac{L_{_{EMHD}}}{L_{_{MHD}}}\right)\ll 1,
\end{equation} 
must hold since the characteristic quantities in EMHD must be much smaller than $d_i$ and those of RMHD much larger than $d_i$. 

\subsection{Linear dispersion relations}\label{linear}
We now focus on the collisionless regimes, $S^{-1}=S^{-1}_{_{EMHD}}=0$; we will not consider viscous effects, whose role in MHD has been clarified recently by \emph{Tenerani et al. (2015)}
\cite{viscous}. In addition, to further simplify the analysis, we start by setting $\rho_s=0$ in Eqs.(\ref{eq:OhmRMHD})-(\ref{eq:EMHD_U}). Because of
the fact that both the (squared) electron skin depth and the Lundquist number weigh non ideal terms in Ohm's law which allow magnetic lines to reconnect (Appendix \ref{App:model_equations}), and of other similarities which will be later discussed,
let us introduce for future use the notations\, $\varepsilon_d \equiv d_e^2$\, and\,
$\varepsilon_{_S}\equiv S^{-1}$. Then, after re-scaling, we will write
\begin{equation}\label{eq:rescaling_epsilon}
\varepsilon_d^*=\varepsilon_d\left(\frac{a}{L}\right)^2,\qquad\qquad
\varepsilon_{_S}^*=\varepsilon_{_S}\left(\frac{a}{L}\right).
\end{equation}

 After linearizing Eqs.(\ref{eq:OhmRMHD})-(\ref{eq:RMHD_U}) around an equilibrium  $\psi_0(x/a)$ with  perturbations of the form $\sim e^{iky+\gamma t}$, analytic approximations to the
dispersion relations in both  RMHD and EMHD may be obtained  by applying the boundary layer technique, as first shown by \emph{Furth et al. (1963)} \cite{FKR}.
 
Here we summarize the results valid in the two asymptotic regimes called large (LD) and small (SD) $\Delta'$, which respectively correspond to the  internal kink and  constant-$\psi$ orderings \citep{Ara}. In RMHD such regimes are respectively  defined  by the conditions $\Delta'\delta > 1$ (LD) and $\Delta'\delta < 1$ (SD), where $\delta$ is the characteristic reconnection layer width. The inertial RMHD tearing dispersion relations become (see e.g. \citep{Porcelli}): 
\begin{equation}\label{eq:disp_RMHD}
\mbox{RMHD }\left\{\begin{array}{c}
\displaystyle{\gamma_{_{LD}}\tau_{_A}=k d_e}
  \\
\displaystyle{\gamma_{_{SD}}\tau_{_A}=  (C_1 \Delta')^{2} kd_e^3}
  \\
\end{array}\right., 
\end{equation}
where $C_1\equiv \Gamma(1/4)/(2\pi\Gamma(3/4))\simeq 0.4709$. 

In EMHD, where the LD limit corresponds more properly to the condition $\gamma_{LD}/k\sim$\,\emph{constant}, we consider the dispersion relations
\begin{equation}\label{eq:disp_EMHD}
\mbox{EMHD }\left\{\begin{array}{c}
\gamma_{_{LD}}\tau_{_W}=\displaystyle{
C_2k d_e^\frac{2}{3}}   \\
\gamma_{_{SD}}\tau_{_W}=\displaystyle{ \left( C_1\Delta'\right)^2d_e^2} \\
\end{array}\right.,
\end{equation}
where $C_2\equiv(2\Gamma^4(3/4))^{-1/3}\simeq 0.6053$. 
The $\gamma_{_{LD}}$  growth rate above, which is the one evaluated  
by \emph{Attico et al. (2000)} \cite{Attico}  starting from an equilibrium given by $\psi_0(x)=x/a$ for $-a<x<a$ and $\psi_0(x)=1$ for $|x|\geq a$, has been assumed as the prototype for the more general ``LD'' EMHD dispersion relation for a generic sheared, even, $\psi_0(x)$ profile. The reason is that  this is the only available formula obtained for this wavelength regime, and,  with the same equilibrium, the general $\gamma_{_{SD}}$ dispersion relation first computed in \citep{Bulanov} and quoted in Eq.(\ref{eq:disp_EMHD}), was exactly recovered.

\begin{figure}[h]
\includegraphics[width=18pc]{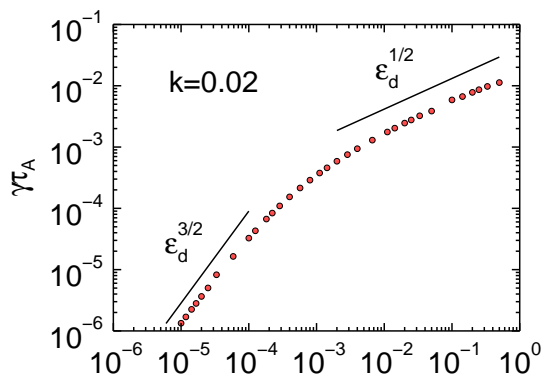}
 \caption{Scaling of $\gamma(\tilde{k})\tau_{_A}$ in the RMHD regime as a function of $d_e^2$ for a fixed $\tilde{k}$. At the increase (decrease)  of $d_e$ the small (large) $\Delta'$ regime is progressively entered. Here $\tilde{k}=k_{_M}$ for $d_e^2\simeq 2\times 10^{-3}$ (lengths in units of $a$). }
 \label{Fig_LD_SD}
 \end{figure}

For illustrative purposes in Fig.\ref{Fig_LD_SD} we show the scaling of the growth rate of a given unstable mode $\tilde{k}$ as a function of $\varepsilon_d$  in the RMHD regime.  Notice that the whole range of regimes  from SD to LD is spanned while varying the value of $d_e$ at given $k$. Indeed, since $\delta=\delta(k,\varepsilon_d)$, an interval in the $\varepsilon_d$ parameter space such that 
$\Delta'(\tilde{k})\delta(\tilde{k},\varepsilon_d)$ is  smaller (SD),  equal ($\gamma_{_M}$, see Sec.\ref{Results_collisionless}), or greater (LD) than unity,  always exists.

As a comment,  note that almost ideal growth rates (saturating at $(\gamma_{_{LD}}^{_{EMHD}})^*\simeq 0.25 (\tau_{_W}^*)^{-1}$) were observed in numerical integrations of the EMHD linear system at $ 0.1\lesssim d_e < 1$ \citep{EMHD}  for $L/a=2\pi$ and  $k^*=k=1$. Such large values of  $d_e$ are not unreasonable in the collisionless EMHD regime, because of the constraint $d_e\ll a\ll d_i$ (now in dimensional units), which must be fulfilled by the equilibrium shear length. With such large values of the reconnection parameter, we are outside the realm of the asymptotic/boundary layer analysis, but for EMHD this is to be expected, since characteristic EMHD scale lengths must satify $\ell$ fulfill $d_e \ll \ell$, or, given that $d_i/d_e\simeq 42Z$ for an ion charge $Z$, $d_e\ll \ell \ll 42 d_e Z$. 
Similarly large growth rates are found in strongly resistive RMHD regimes $S^{-1}\gtrsim 0.01$, though these are normally of little interest. Discrepancies with analytical estimations from Eqs.(7)-(8), suggest that at $\varepsilon_d \sim 0.01$ or equivalently 
$\varepsilon_S \sim 0.01$ the boundary layer approach to the linear tearing breaks down.

\section{Results}\label{Results}

\subsection{Transition to the inertial ideal regime}\label{Results_collisionless}
 When $L/a\gg 1$, say, $L/a\gtrsim 20$ \citep{Velli}, we can search for the fastest unstable  mode $k_{_M}$ with corresponding growth rate $\gamma_{_M}$. As noticed by \emph{Battacharjee et al. (2009)} \cite{Batta}, the latter can be estimated by imposing the condition $\gamma_{_{LD}}(k_{_M})=\gamma_{_{SD}}(k_{_M})\equiv\gamma_{_M}$.  Approximating  $\Delta'(k_{_M})\simeq Kk_{_M}^{-p}$ where $K$ is a constant,  from Eqs.(\ref{eq:disp_RMHD}) and  Eqs.(\ref{eq:disp_EMHD}) we can estimate (see also \citep{equilibri}):
\begin{equation}\label{eq:max_RMHD}
\mbox{RMHD }\left\{\begin{array}{c}
\displaystyle{ k_{_M}\simeq \left( KC_1\right)^{\frac{1}{p}} d_{e}^{\frac{1}{p}}} \\
 \displaystyle{\gamma_{_M}\tau_{A}\simeq \left( KC_1 \right)  d_e^{\frac{1+p}{p}}},\\ 
\end{array}\right.
\end{equation}
\begin{equation}\label{eq:max_EMHD}
\mbox{EMHD }\left\{\begin{array}{c}
\displaystyle{ k_{_M}\simeq \left(\frac{K^2C_1^2}{C_2}\right)^{\frac{1}{1+2p}}
d_e^{\frac{4}{3(1+2p)}}} \\
 \displaystyle{ \gamma_{_M}\tau_{W}\simeq
 \left(KC_1C_2^p\right)^{\frac{2}{1+2p}} d_e^{\frac{2}{3}\frac{3+2p}{1+2p}}}.\\ 
\end{array}\right.
\end{equation} 
 
Let us now apply the re-scaling argument to evaluate, from Eqs.(\ref{eq:max_RMHD}-\ref{eq:max_EMHD}) and from the definitions of $\tau_{_A}$ and $\tau_{_W}$, the scaling of the most unstable mode when lengths are normalized to $L$. Neglecting the numerical coefficients in the parentheses of Eqs.(\ref{eq:max_RMHD}-\ref{eq:max_EMHD}) we find in RMHD, 
 \begin{equation}\label{eq:RMHD_max_ideal}
\displaystyle{ k_{_M}^*\simeq 
(\varepsilon_d^*)^{\frac{1}{2p}}\left(\frac{L}{a}\right)^{\frac{1}{p}}}\, 
\quad
\displaystyle{ \gamma_{_M}\tau_{_A}^*\simeq 
(\varepsilon_d^*)^{\frac{1+p}{2p}}\left(\frac{L}{a}\right)^{\frac{1+2p}{p}}}\,,
\end{equation}
and in EMHD
\begin{equation}\label{eq:EMHD_max_ideal}
\displaystyle{ k_{_M}^*\simeq 
(\varepsilon_d^*)^{\frac{2}{3+6p}}\left(\frac{L}{a}\right)^{\frac{4}{3+6p}}}\, 
\quad
\displaystyle{ \gamma_{_M}\tau_{_W}^*\simeq 
(\varepsilon_d^*)^{\frac{3+2p}{3+6p}}
\left(\frac{L}{a}\right)^{\frac{12+16p}{3+6p}}}\, .
\end{equation}
 In the RMHD regime it is easy to verify from the analytical estimates $\delta_{_{LD}}\sim d_e$ and $\delta_{_{SD}}\sim \Delta'd_e^2$  (see e.g. \citep{Ottaviani1})  that the fastest growing mode satisfies the condition  $\Delta'(k_{_M})\delta(k_{_M})\sim 1$.  The characteristic width of the reconnection layer for the most unstable RMHD mode therefore becomes 
\begin{equation}\label{eq:RMHD_delta}
\displaystyle{ \delta_{_M}\simeq d_e},
\end{equation}
which, after rescaling, reads $\displaystyle{ \delta_{_M}^*\simeq (\varepsilon_d^*)^{\frac{1}{2}} }$.

The condition for ``ideal'' tearing is set by searching for the value of $\alpha$ such that when $a/L\sim (\varepsilon_d^*)^{\alpha}$ with $\alpha>0$  $\gamma_{_M}^*$ becomes independent of $\varepsilon_d^*=\varepsilon_d a^2/L^2$. 
Imposing this, we find the exponent $\alpha$ both in RMHD and EMHD, respectively,
\begin{equation}\label{eq:alpha}
 \alpha_d^{_{RMHD}}=\frac{1+p}{2+4p},\qquad\quad 
\alpha_d^{_{EMHD}}=\frac{3+2p}{12+16p}\, .
\end{equation}
In particular, for a Harris-pinch equilibrium, which has  $p=1$, we find
 \begin{equation}\label{eq:alpha_Harris}
 \alpha_d^{_{RMHD}}=\frac{1}{3}, \qquad\quad\alpha_d^{_{EMHD}}=\frac{5}{28}
\simeq 0.1786.\end{equation}
 
\begin{figure}
 \noindent\includegraphics[width=20pc]{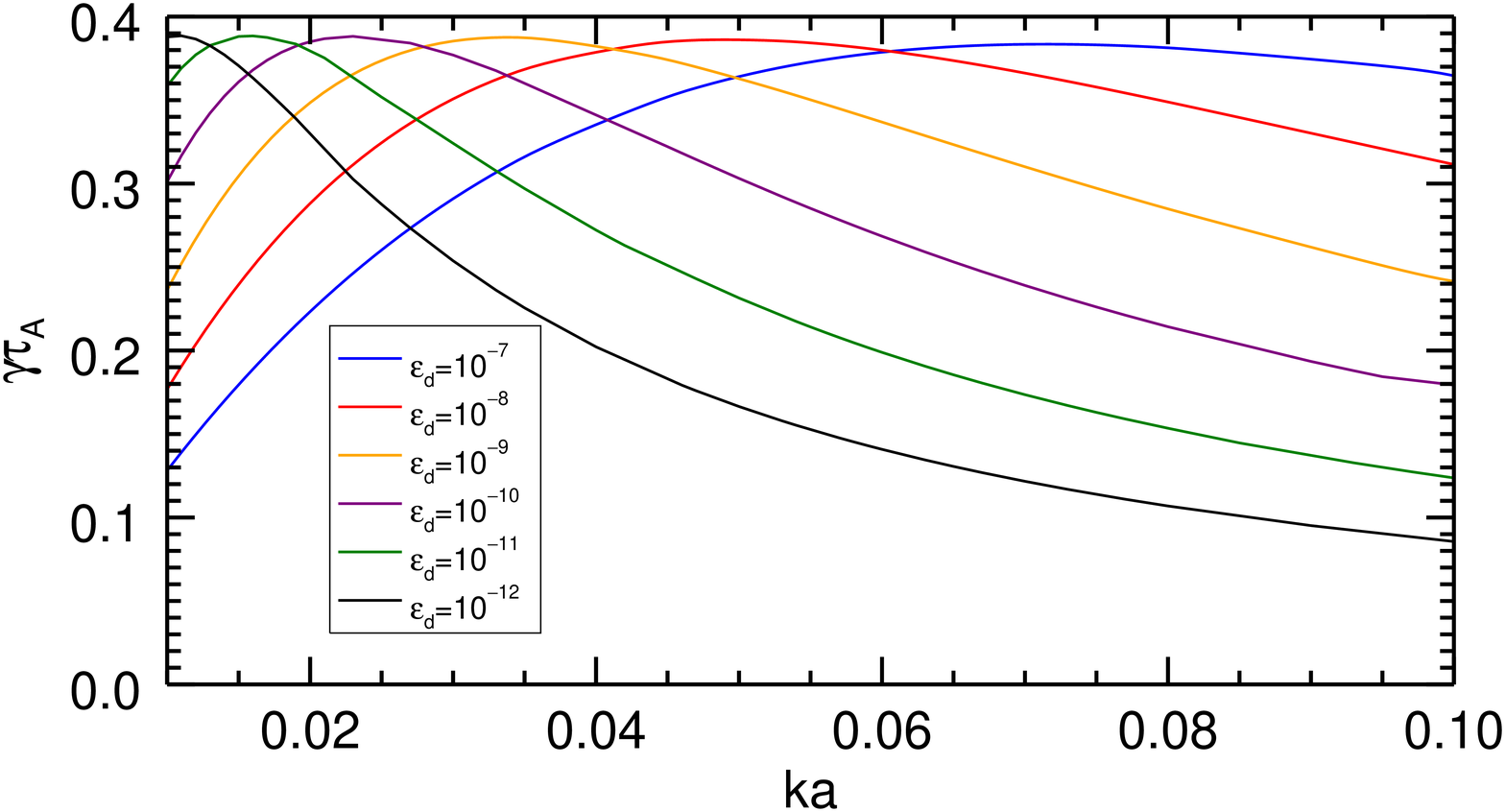}
\includegraphics[width=20pc]{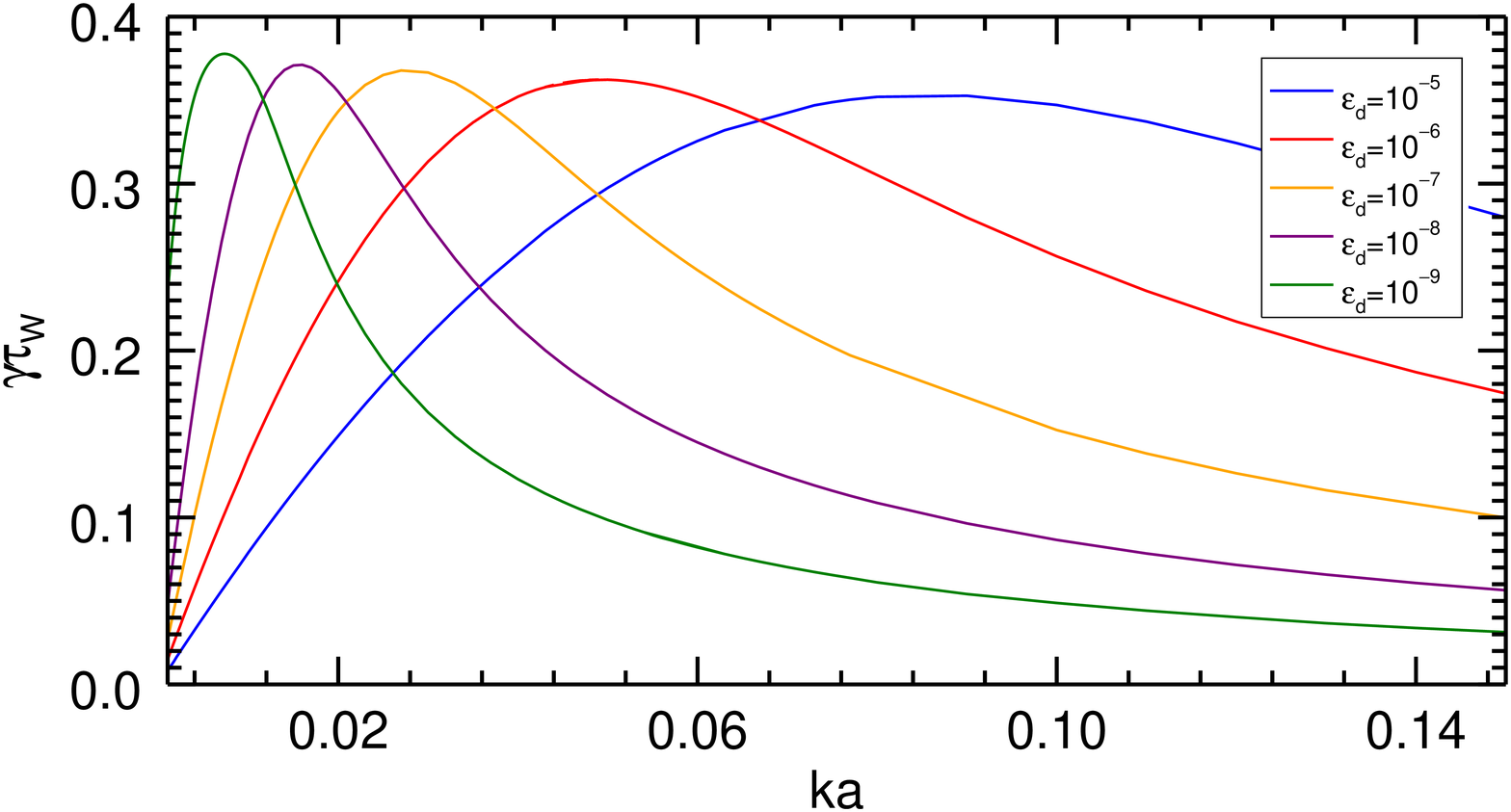}
 \caption{RMHD (left frame) and EMHD (right frame) dispersion relations $\gamma^*=\gamma^*({k}^*,\varepsilon_d^*)$, computed  for different values of $d_e$ and represented as functions of $k^*a^*$. For each curve an aspect ratio was chosen, satisfying  the threshold condition for a Harris-pinch equilibrium, $a/L=(\varepsilon_d^*)^{1/3}$  in RMHD and $a/L=(\varepsilon_d^*)^{3/16}$ in EMHD. The maximum growth rate on each curve is independent from $d_e$ and of order unity with respect to the characteristic time: $\gamma_{_M}^*\tau_{_A}^*\simeq 0.39 $ in RMHD and $\gamma_{_M}^* \tau_{_W}^*\simeq 0.37$ in EMHD.}
 \label{Fig_ideal}
 \end{figure}

A set of curves $\gamma({k})$ for different values of $d_e$ along the RMHD threshold condition $a/L=(\varepsilon_d^*)^{1/3}$ is plotted  in Fig.\ref{Fig_ideal}a, while the corresponding graph for the EMHD regime is in Fig.\ref{Fig_ideal}b. 
The independence of $\gamma_{_M}^*$ from $d_e$ and its value of order unity, namely $\simeq 0.39 (\tau_{_A}^*)^{-1}$ in RMHD and $\simeq 0.37 (\tau_{_W}^*)^{-1}$ in EMHD, is evidenced in both regimes. Referring to the example of the Harris-pinch profile and assuming for the EMHD the numerical  threshold condition $a/L=(\varepsilon_d^*)^{\frac{3}{16}}$, we then deduce the scalings of the  threshold current sheet widths $a$ with respect to $d_e$, which will be discussed in Sec.\ref{summary}: 
\begin{equation}\label{eq:a_vs_de}
\left( \frac{a}{d_e}\right)_{_{RMHD}}= \left( \frac{L}{d_e}\right)^{\frac{1}{3}},
\qquad \left( \frac{a}{d_e} \right)_{_{EMHD}}= \left( \frac{L}{d_e}\right)^{\frac{5}{8}}.
\end{equation}

\subsection{Kinetic effects in the transition to the inertial ideal tearing: FLR corrections}\label{Results_FLR}
We now briefly consider the role played by  other kinetic effects important at small spatial scales  $a\ll L$ where the transition to ``ideal'' tearing takes place. Since ion-ion viscosity effects have been already discussed by \emph{Tenerani et al. (2015)} \cite{viscous} we focus on  FLR effects, which enter in our set of equations through the so-called gyrofluid corrections, an example of which is provided by the $\rho_s$ term in Eqs.(\ref{eq:OhmRMHD}-\ref{eq:RMHD_U}).

At small scales $\ell\ll L$ the fluid description formally breaks down, but it has been shown that gyrofluid RMHD models capture the essential physics of gyrokinetic reconnection \citep{Zacharias}. A good agreement between our collisionless RMHD equations at $\rho_s\neq 0$ and  a drift-kinetic model for magnetic reconnection was already pointed out  \cite{Peg3}. The ion-sound Larmor radius was shown to increase the inertia-driven tearing reconnection rate both linearly \citep{Peg1,Peg2,Porcelli} and nonlinearly \citep{Ottaviani1,Cafaro,Grasso1,PPCF,Comisso1,Comisso2,Hirota}. Notice that the RMHD equations have been extended to include also ion FLR effects,  $\rho_i\equiv v^{i}_{th}/\Omega_{ci}$ ($v_{th}^i$ being ion thermal velocity), related to the ion-sound Larmor radius by $\rho_s^2=\rho_i^2T_e/T_i$.  These effects are usually introduced in RMHD equations, notably in Eq.(\ref{eq:RMHD_U}), by making some closure assumption on the ion kinetic response obtained from the transport equation. Different models are then available, but also those whose different Hamiltonian properties were compared by \emph{Welbroeck et al. (2009)} \cite{Waelbroeck}, were shown to provide numerical results in a remarkably good agreement \citep{Grasso2,PPCF}. Also notice that the isothermal assumption behind the definition of $\rho_s$ and $\rho_i$ has been shown to be in good agreement with the numerical results from gyrokinetic models for electrons, during the whole linear reconnection stage \citep{Perona}. Interestingly, in a certain parameter range, the theoretically predicted scalings of tearing modes \citep{Peg1,Porcelli}  display a symmetric dependence on the two FLR effects, as the latter enter in the dispersion relation as powers of $\rho_\tau^2=\rho_s^2+\rho_i^2$. Even if appreciable discrepancies from these predictions are seen as the ratio $\rho_\tau^2/d_e^2$ increases \citep{PPCF}, at $\Delta'd_e\gg \min{[1, (d_e/\rho_\tau)^{1/3}]}$ a good agreement is found \citep{Comisso1}. 

In the regime $\rho_\tau^2\gg d_e^2$,  \emph{Comisso et al. (2013)} \cite{Comisso2} recently pointed out the existence of a maximum growth rate in the continuum spectrum limit (i.e. continuous $k$) of unstable tearing modes, corresponding, in our notation, to $k_{_M}$. The generalization of the result they obtained for the Harris-pinch case to generic equilibria, is obtained as described in Sec.\ref{Results_collisionless}, by  starting from their Eqs.(26)-(27) instead of our Eqs.(\ref{eq:disp_RMHD}). We find
\begin{equation}\label{eq:FLR_max}
\displaystyle{
k_{_M}\simeq 
d_e^{\frac{2}{3p}}\rho_\tau^{\frac{1}{3p}}},
\qquad
\displaystyle{ \gamma_{_M}\tau_{_A}\simeq 
d_e^{\frac{2+p}{3p}}\rho_\tau^{\frac{1+2p}{3p}}}\, 
.\end{equation}
Then, applying the rescaling arguments, we obtain $\gamma_{_M}^{_{FLR}}\tau_{_A}^*\sim O(1)$ when 
\begin{equation}\label{eq:FLR_threshold}
\frac{a}{L}\sim (\varepsilon_d^*)^{\frac{2+p}{6+12p}}
(\rho_\tau^*)^{\frac{1}{3}}\,.\end{equation}
 We then see that, depending on the value of the ratio $d_e/\rho_\tau$, the inclusion of FLR corrections may imply an even larger critical aspect ratio for the transition to  ``ideal'' tearing, with respect to the cold-plasma limit. Indeed, if we now assume $\rho_\tau\simeq A d_e$ and we compare the threshold condition of
 Eq.(\ref{eq:FLR_threshold}) with that of Eq.(\ref{eq:alpha}) for the RMHD, we see that, with obvious notation, the two  are related through $(a/L)_{_{FLR}}\sim A^{1/3}(a/L)_{_{RMHD}}$. Since usually $A>1$ (e.g., tipically  $A\sim 10$ in tokamak plasmas and it may be even larger in the magnetosphere  this implies a broadening of the ideally unstable current sheet with respect to the cold plasma case, when kinetic effects are taken in account.   

\section{Discussion}\label{Discussion}

\subsection{Collisionless ideal tearing in space, solar and laboratory plasmas}\label{inertia}
 In order to discuss the relevance of electron inertia and resistivity in various natural and laboratory environments where low-collision reconnection occurs,  different plasma parameters, including $ \varepsilon_S^*$ and $\varepsilon_d^* $,  are  shown in Table \ref{Table_1}. 

We recall that the condition for purely collisionless reconnection ($S^{-1}=0$) is given by  $\gamma_d\tau_{_A}\varepsilon_d\gg \varepsilon_{_S}$, with $\gamma_d$ reconnection rate of the sheer inertia-driven regime. Notice that this condition becomes less critical when approaching the ideal regime ($a/L\ll 1$), where $\gamma_d^*\rightarrow 1$  because of the rescaling, which in RMHD implies $\varepsilon_{_S}/\varepsilon_d=(a/L)\varepsilon_{_S}^*/\varepsilon_d^* $ 
(Eqs.(\ref{eq:rescaling_epsilon})): if $\varepsilon_{_S}^*/\varepsilon_d^*\ll 1$, then we can assume the IT model applied to large aspect-ratio current sheets as essentially inertia-driven.  This means, for example, that the magnetotail is in an essentially inertia-dominated tearing regime. On the other hand, fusion devices, for which $a\simeq L$, may operate in conditions in which the resistive contribution to tearing reconnection is not negligible even if $\varepsilon_{_S}^*/\varepsilon_d^*\sim 
\varepsilon_{_S}/\varepsilon_d\sim 10^{-2}-10^{-3}$, because of the smallness of $\gamma_d\tau_{_A}^*$, which remains of the same order of $\gamma_d\tau_{_A}\ll 1 $.  

\begin{figure}
\noindent\includegraphics[width=20pc]{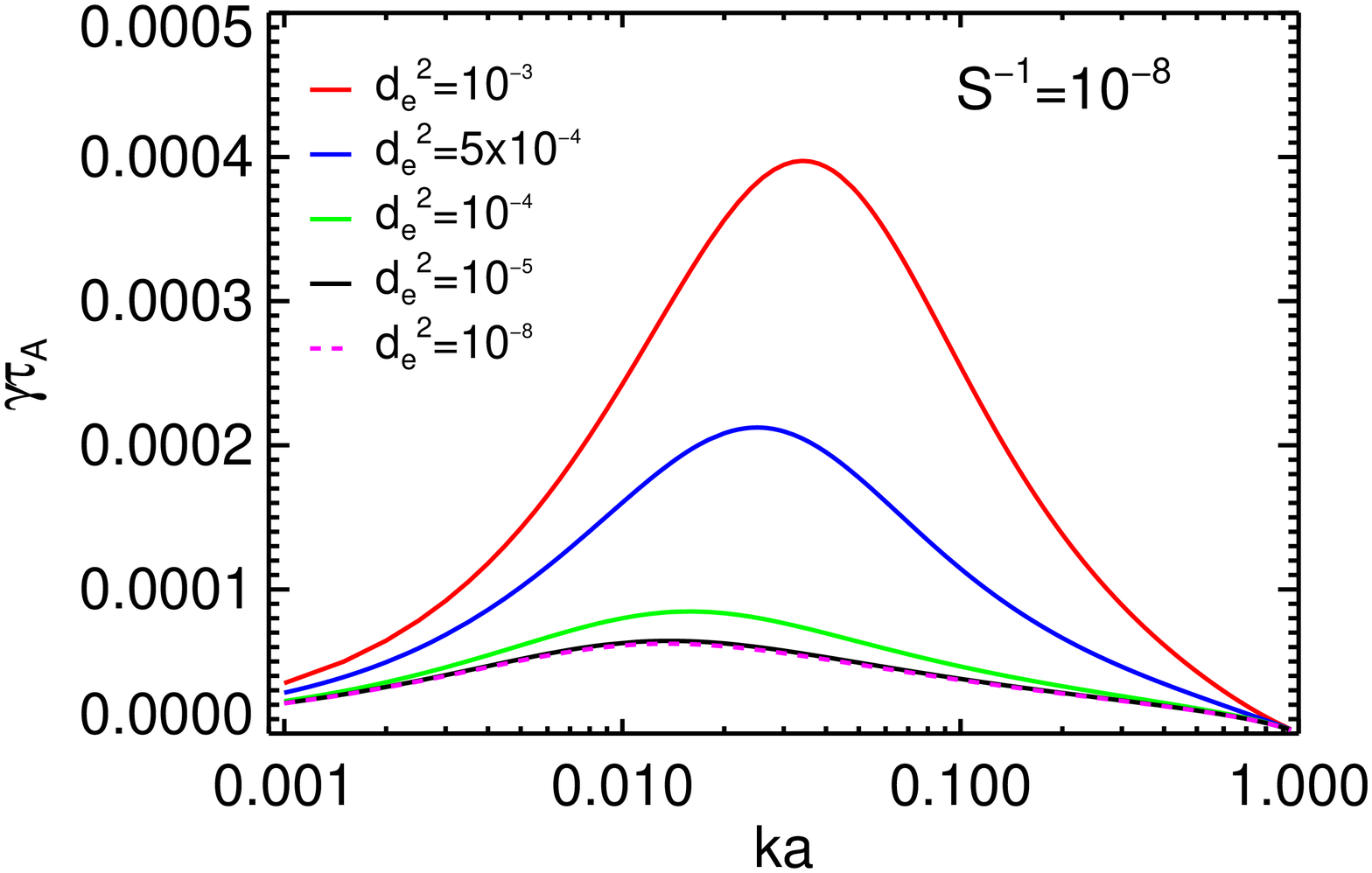}
\noindent\includegraphics[width=20pc]{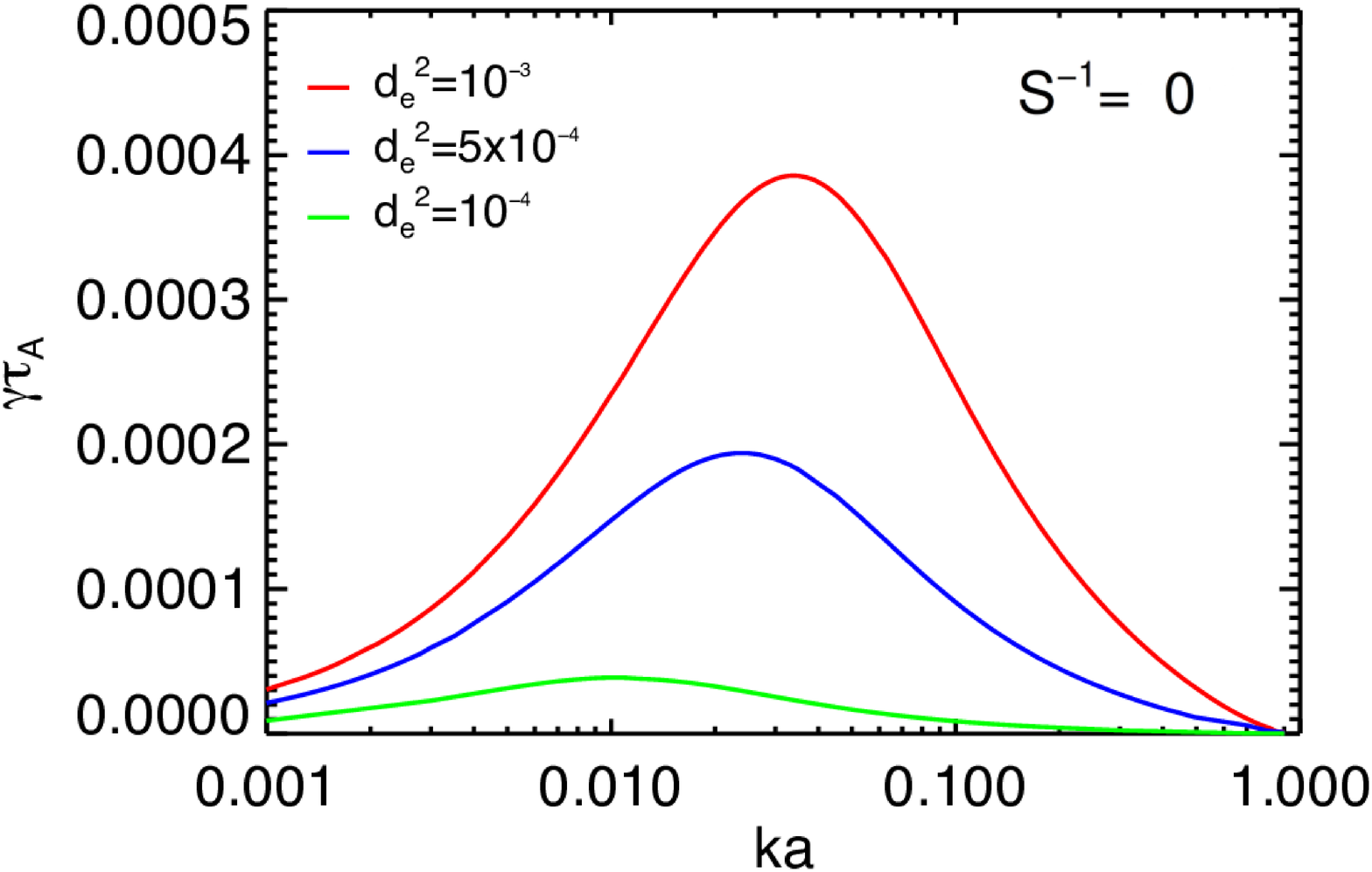}
 \caption{Dispersion relations $\gamma$ vs. $k$ for different values of $d_e$ and for $S^{-1}=10^{-8}$ (upper panel) and $S^{-1}=0$ (lower panel). Some orders of magnitude of separation between the purely inertial and purely resistive growth rates (tipically about $3-4$, at least) are needed in order for $\varepsilon_S$ to be really negligible. }
 \label{de_vs_S}
 \end{figure}

This is because electron inertia, $\varepsilon_d$, enters in the  the dispersion relation of tearing modes with a less favorable scaling with respect to resistivity, $\varepsilon_{_S}$. For practical purposes, at $a/L\sim 1$ the collisionless regime is essentially inertial if $\varepsilon_{_S}$ is sufficiently small ($\varepsilon_{_S}^{-1}\lesssim 10^{-8}$) and $\varepsilon_d$ is at least $3-5$ orders of magnitude larger than $\varepsilon_{_S}$. 
 The case in which  the inertial $\gamma_d$ may dominate over the resistive $\gamma_{_S}$, is exemplified in Figs.\ref{de_vs_S}, where  some examples of the inertial-resistive growth rate are represented, for which only an implicit analytical expression for $\gamma$ is available  (see e.g. Eq.(16) of \citep{Ottaviani1}). The dispersion relations displayed are obtained by numerical integration of the linearized Eqs.(\ref{eq:OhmRMHD}-\ref{eq:RMHD_U}).  These examples show that no appreciable differences in the inertial-resistive growth rates with $\varepsilon_{_S}=10^{-8}$ are observed between  $\varepsilon_d=10^{-8}$ and $\varepsilon_d=10^{-5}$.  At higher values of $S^{-1}$, both the inertial and the resistive contributions to the  inertial-resistive growth rate become appreciable, and for $S^{-1}\gtrsim 10^{-6}$ the resistive contribution to the growth rate is relevant even for $d_e$ approaching unity.

\subsection{Ideal tearing and stability of steady-state reconnecting current sheets in the collisionless regime}\label{steady}
 Both in  MHD \citep{Wesson} and in EMHD \citep{Bulanov,Avinash}, the reconnection rate of a steady state current sheet has been evaluated in the collisionless regime, as a generalization of the classic Sweet-Parker configuration. In both\footnote{Notice that Wesson's result \citep{Wesson} was specialized to a geometric configuration corresponding to the $m=1$ mode in a tokamak, but his reasoning is easily adapted to the standard planar sheet configuration.} cases the same scaling in $\varepsilon_d$ of the stationary Sweet-Parker-like reconnection rate $\gamma_{_{SP}}$  was obtained with respect to the respective normalization times, 
$(\tau_{_{SP}}^{_{EMHD}})^{-1}\tau_{_W}^*\sim (\varepsilon_d^*)^{1/2} $ and $(\tau_{_{SP}}^{_{RMHD}})^{-1}\tau_{_A}^*\sim (\varepsilon_d^*)^{1/2} $. This  implies that both in collisionless RMHD and EMHD, the aspect ratio scaling of a steady current sheet of length $L$ is $(a/L)_{_{SP}}\sim (\varepsilon_d^*)^{1/2}$. By comparing the scaling of this ratio with the threshold conditions for the onset of ``ideal'' tearing (Eqs.(\ref{eq:alpha})) the same qualititative behavior, though with different scalings, is evidenced in both RMHD and EMHD. In RMHD the width of the steady reconnecting layer  corresponds to  a much thinner  current sheet than that which is unstable to  ideal tearing: at a given length $L$, the collisionless Sweet-Parker sheet width, $a_{_{SP}}$, is related to the ideal-tearing unstable one, $a_{_{IT}}$, by the relation $a_{_{SP}}^*\simeq (a_{_{IT}}^*)^{(1+2p)/(1+p)}$. Using the same reasoning  we can estimate from Eq.(\ref{eq:alpha})  $a_{_{SP}}^*\simeq (a_{_{IT}}^*)^{(6+8p)/(3+2p)}$ for EMHD. If we now neglect the effect  of the flow along the neutral line on the growth rate (cfr. also \citep{id_tearing} for why flows may be neglected), this means that both in  RMHD  and EMHD a collisionless Sweet-Parker-type current sheet is always unstable on ideal time scales. 

\subsection{``Secondary'' ideal tearing and ``explosive reconnection''}\label{secondary}

 The rescaling argument at the basis of ideal tearing may thus provide a fairly general paradigm to describe explosive growth rate increases observed in the nonlinear stage of simulations of reconnection at $L/a$ not much larger than unity \citep{Ali,Biancalani,Yu}, when an $X$-point collapses into two $Y$-points and the current sheet between the two becomes tearing unstable, eventually leading to the so-called plasmoid-chain instability. During this stage, even before an ideal growth rate is achieved,  a secondary growth rate  may be measured, which is arbitrarily large (possibly up to the inverse macroscopic time scale, in the ideal tearing limit).

  Let be $L_{_Y}$ the length and $a_{_Y}$ the width of a secondary current-sheet between  two $Y$-points, generated in the nonlinear stage of the tearing of a current sheet with aspect ratio $a/L$. 
Focusing on the dynamics of this secondary current sheet, the CT growth rates  would refer lengths to $a_{_Y}$, whereas  we  now need to label with ``$\tilde{...}$'' the quantities normalized to  $L_{_Y}$, since the latter plays the role of macroscopic length for the secondary dynamics (cfr. Sec.\ref{Ideal_tearing}).  Even when we consider a primary tearing mode with $L/a\gtrsim 1$, the secondary current sheet develops with a much smaller thickness (corresponding to the singular layer thickness of the original tearing instability)  so that we can assume the most unstable tearing mode to be destabilized:
accounting for FLR effects, the re-normalized, most unstable, 
tearing mode growth rate on the secondary current sheet  (cfr. Sec.\ref{Results_FLR}) therefore becomes dependent from the ratio $L_{_Y}/a_{_Y}$,
\begin{equation}\label{eq:secondary_FLR}
{\gamma}_{_M}^{_{FLR}}\tilde{\tau}_{_A}\sim \tilde{\varepsilon}_d^{\frac{2+p}{6p}}\tilde{\rho}_{\tau}^{\frac{1+2p}{3p}}
 \left(\frac{L_{_Y}}{a_{_Y}}\right)^{\frac{1+2p}{p}}.\end{equation}
Analogously, we can rewrite 
 the correspective for the resistive inviscid and viscous, high-Prandtl number RMHD regimes, respectively discussed in \citep{id_tearing} and \citep{viscous,cascade},
\begin{equation}\label{eq:secondary_S}
{\gamma}_{_M}^{_{_{res}}}\tilde{\tau}_{_A}
\sim \tilde{\varepsilon}_{_S}^{\frac{1+p}{1+3p}}
 \left(\frac{L_{_Y}}{a_{_Y}}\right)^{\frac{2+4p}{1+3p}},
\end{equation}
\begin{equation}\label{eq:secondary_S_visc}
{\gamma}_{_M}^{_{_{visc}}}\tilde{\tau}_{_A}\sim 
\tilde{\varepsilon}_{_S}^{\frac{1+2p}{1+3p}}
\tilde{R}^{\frac{p}{1+3p}}
 \left(\frac{L_{_Y}}{a_{_Y}}\right)^{2}.
\end{equation}

The occurrence of a secondary, ideal tearing mode developing as a consequence of a primary tearing in a large aspect ratio current sheet in the resitive RMHD regime was first numerically evidenced by \emph{Landi et al. (2015)} \cite{Landi} and discussed in depth in \citep{cascade}.

Let us now  focus on a primary tearing mode of a small aspect ratio current sheet, assuming for simplicity  $a\sim L$. In this case the primary reconnection rate can not be estimated with that of the most unstable mode $\gamma_{_M}$ but the specific LD or SD regime in which the unstable wave-number falls must be taken in account, instead.    Let us now compare such primary reconnection rate to the secondary one, as estimated from Eqs.(\ref{eq:secondary_FLR}-\ref{eq:secondary_S_visc}). We immediately recognize that,  even before the ideal tearing threshold is reached,  the re-scaling argument predicts an increase  in the growth rate, \emph{measured with respect to the primary mode macroscopic scale  $L$}, by some positive power of $(L_{_Y}/a_{_Y})>1$ times some positive power of $(L/L_{_Y})>1$. Comparing Eqs.(\ref{eq:secondary_FLR}-\ref{eq:secondary_S_visc}) we see that, for equal equilibrium profiles (same $p\geq 1$), such an increase is relatively more important in the inertia driven-FLR regime.

To fix the ideas with a quantitative example, consider the RMHD-FLR regime supposing the primary reconnection to develop on a current sheet described by the equilibrium used by \emph{Comisso et al. (2013)} \cite{Comisso2}, assuming an aspect ratio so close to unity that a  single primary mode $m_0$ (i.e. $k_0=2\pi m_0/L$) is excited in the SD, constant-$\psi$ regime, in the whole range of parameters in which $d_e$ and $\rho_\tau$ are varied ($\Delta'\rho_\tau^{\frac{1}{3}}d_e^{\frac{2}{3}} < 1$). The primary tearing mode (see e.g. Eq.(27) of \citep{Comisso2}), once rescaled to $L$,  grows with  $\gamma_I\tau_{_A}^*\simeq k_0^*(\varepsilon_d^*)^\frac{1}{2}\rho_\tau^*(\Delta')^*(L/a)$, with some $(\Delta'(k_0^*))^*$ of order unity. For the secondary mode we may now use  Eq.(\ref{eq:secondary_FLR}).  Assuming for simplicity (but with no loss of generality)  that the secondary current sheet resembles a Harris-pinch profile to specify some value of $p$ (here $p=1$), the secondary growth rate, expressed again in terms of the scale $L$, is given by  Eq.(\ref{eq:secondary_FLR}) opportunely rescaled, $\gamma_{II}\tau_{_A}^*\sim (\varepsilon_d^*)^\frac{1}{2}\rho_\tau^* (L/a_{_Y})^3$. A dominant increase of the reconnection rate is therefore provided by the ratio $L/a_{_Y}\gg 1$. In  particular, in this example we obtain $\gamma_{II}^*/\gamma_{I}^*\sim (a/a_{_Y})(L/a_{_Y})^2(k_0^*\Delta'^*)^{-1}$.

Of course, a more detailed analysis would be required to verify whether the re-scaling argument summarized by Eqs.(\ref{eq:secondary_FLR}-\ref{eq:secondary_S_visc}) and the corresponding  threshold conditions for the ideal tearing  suffice to explain the explosive reconnection regimes observed in the above mentioned numerical studies. However, the qualitative considerations about the scalings provided in Fig.(3) of \citep{Biancalani_EPS} and in Fig.(2) of \citep{Biancalani} seem encouraging. Because of the normalization assumed in these articles, the increase of the growth rates with decreasing plasma $\beta$ implies for the linear growth rate a scaling $\gamma_I\tau_{_A}^* \sim d_e$ and for the nonlinear one a scaling $\gamma_{II}\tau_{_A}^*\sim d_e^0$ at fixed $\rho_s$, thus suggesting  (cfr. Eq.(\ref{eq:FLR_max}) and Eq.(\ref{eq:secondary_FLR}) for $p=1$) that an ideal tearing regime was observed in the nonlinear stage of the simulations discussed by \emph{Biancalani et al. (2012)} \cite{Biancalani}.  Future studies will elucidate whether the explosive reconnection predicted by Eq.(\ref{eq:secondary_FLR}) and that studied in \citep{Biancalani} are effectively  the same phenomenon.  

 To conclude this Section, we finally notice that, provided the ratio $L_{_Y}/a_{_Y}$ is large enough to destabilize a most unstable mode $\gamma_{_M}$ (which is typical for secondary current sheets developed from the collapse of an $X$-point in the resistive regime \citep{cascade}) the measured growth rate would be that of an exponentially growing instability, which in the resistive regime has the same scaling with $S$ as the Sweet-Parker reconnection rate (i.e. $\sim S^{-1/2}$). 

\section{Summary}\label{summary}
We have extended the analysis of \citep{id_tearing} to collisionless regimes, both in RMHD and EMHD, by providing the scaling threshold values $a/L\sim(d_e^2/L^2)^{\alpha}$ at which a current sheet with $L/a\gtrsim 20$ reconnects on the ideal macroscopic times of the model. For the Harris-pinch equilibrium profile the exponents measured after numerical solution of the eigenvalue problem are $\alpha_d^{_{RMHD}}=1/3$ and $\alpha_d^{_{EMHD}}\simeq 3/16$, in excellent agreement with the analytical estimations obtained by starting from the SD and LD dispersion relations.  In RMHD, FLR corrections typically reduce the width of the critical aspect ratio for the transition to ``ideal'' tearing. In the parameter range  $\Delta'd_e\gg \min[1,(d_e/\rho_\tau)^{1/3}]$ and for the Harris-pinch case, such an aspect ratio becomes $(a/L)_{_{FLR}}\sim  (\varepsilon_d^*)^{1/6}(\rho_\tau^*)^{1/3}$, instead of $(a/L)_{_{d}}\sim  (\varepsilon_d^*)^{1/3}$ in the $\rho_s=0$ limit. Since  this implies a broadening of the critical reconnection current layer by a factor $(d_e^*)^{-1/3}(\rho_\tau^*)^{1/3}\sim A^{1/3}$,  when $\rho_\tau\simeq A d_e$  with $A>1$, as it is usually the case, FLR effects are expected to correspondingly  lower the instability threshold.

The collisionless IT model has been applied to discuss the instability of steady collisionless reconnecting current sheets, which, just as in the resistive case, should not be observable as they become unstable to inertia-driven tearing modes on ideal time-scales.  We notice however that the  threshold current sheet to the IT,  found to be thinner in RMHD than in EMHD (Eqs.(\ref{eq:a_vs_de})), leaves the open question of how the Alfv\'enic and whistler-dominated frequency regimes  relate to the Hall-MHD framework, which in principle encompasses both as  two of its limits, opposite one to each other (see Appendix \ref{Ohm_Hall}). 

We have also pointed out the relevance and importance of  inertia-driven vs. resistive reconnection: the condition $S^{-1}\gg d_e^2\gamma$  provides a stringent constraint on when resistivity may be neglected which is often overlooked, for example, when applying Vlasov models of reconnection to tokamak plasmas. 

We have finally discussed how the rescaling argument at the basis of the IT model may explain the ``explosive'' reconnection rate increase observed during the nonlinear stage of primary reconnection events, as secondary elongated current sheets are generated during the collapse of an $X$-point \citep{Ali,Biancalani,Loureiro_2005}.  The IT regime may thus be in principle achieved also during secondary reconnection events involving the thin, elongated current layers nonlinearly generated by classical tearing processes \citep{Ottaviani} or in kinetic turbulence \citep{Servidio}. Notice that large aspect ratio current-layers are generally expected to develop because of the ``exponentiation'' of neighboring magnetic field lines \citep{Boozer_space}), and evidence of such exponential thinning of current sheets was recently provided, in the coronal heating context, by the numerical 3D simulations of \cite{Rappazzo}. This model provides therefore  a promising key to interpretate reconnection rates, which both in laboratory and astrophysics are observed to be orders of magnitude faster than what is predicted by the CT theory. The simplicity of the rescaling argument at the basis of the IT model should not betray its non trivial reach. The dominant trend of recent research on magnetic reconnection, aiming at predicting almost ideal reconnection rates, focuses indeed on the role played by kinetic processes and  secondary instabilities,  whereas the model first considered by \cite{id_tearing} has the appealing feature of relying on simple and well known results. 

%
%

\appendix{}

\section{Discussion of the model equations}\label{App:model_equations}

Eqs.(\ref{eq:OhmRMHD}-\ref{eq:EMHD_U})  are derived with different approximations from the 
electron and ion momentum equations, which we write here below, again non-dimensionalized using  $a$ and $\tau_{_A}$ (and the electric field normalized to a fraction $V_{_A}/c$ of the reference magnetic field):
\begin{equation}\label{eq:electron}
d_e^2
\left( \frac{\partial {\bm u}_e}{\partial t}+{\bm u}_e
\cdot{\bm\nabla}{\bm u}_e\right)=
-d_i\left({\bm E}+{\bf u}_e\times {\bm B} 
- \frac{\bm J}{S}\right)-\rho_s^2\frac{{\bm\nabla}\cdot{\bm\Pi}_e}{n_e} 
 \end{equation}

\begin{equation}\label{eq:ion}
d_i^2
\left(\frac{\partial {\bm u}_i}{\partial t}+{\bm u}_i
\cdot{\bm\nabla}{\bm u}_i\right)=
d_i\left({\bm E}+{\bf u}_i\times {\bm B}
- \frac{\bm J}{S}\right)-{\rho_s^2}\frac{{\bm\nabla}\cdot{\bm\Pi}_i}{n_i} 
\end{equation}
Here the kinetic pressure has been normalized to a reference value $P_0$ for the electron plasma pressure. This explains the weight $\rho_s^2 $ in front of the ion pressure force in Eq.(\ref{eq:ion}), even though the ion thermal Larmor radius is  $\rho_i= (T_i/T_e)^{1/2}\rho_s$.
As discussed in  \citep{MPLB} for the purely collisionless regime, Eqs.(\ref{eq:OhmRMHD}-\ref{eq:EMHD_U}) may be indeed obtained, under appropriate approximations and closures for the pressure tensors (and after re-normalization to $ \tau_{{_W}}$ for the EMHD equations), from Eqs.(\ref{eq:electron}-\ref{eq:ion}) coupled with Maxwell's equations using quasi-neutrality, $n_e=n_i$. Such an approach is essentially the one via which electron inertia effects were first included in reconnection models in the full MHD \citep{Coppi1,Coppi2}) and RMHD  frameworks \citep{Schep}. Within this approach, inclusion of resistive diffusion $S^{-1}$ is straightforward, and the perpendicular ion-ion viscosity too can be retained in the form given in Eq.(\ref{eq:RMHD_U}) if the hypothesis of a strong guide field is also assumed (for a recent discussion see \citep{viscous}).  Derivation of the EMHD equations follows simply from Eqs.(\ref{eq:OhmEMHD})-(\ref{eq:EMHD_U}), since ion dynamics  is  completely neglected \citep{Kingsep}.

It can be  verified that both Eq.(\ref{eq:OhmRMHD}) and Eq.(\ref{eq:OhmEMHD}) represent the $z$-component of electron momentum equation (Eq.\ref{eq:electron}) in the RMHD and EMHD regime respectively, $\psi$ and $-\nabla^2\psi$ expressing  the $z$  component of the vector potential ${\bm A}$ and of the electron current density ${\bm J}$. \\
In RMHD, the $\rho_s^2$ contribution on the r.h.s. of Eq.(\ref{eq:OhmRMHD})  expresses thermal effects related to electron compressibility along the magnetic field lines (see e.g. \citep{Kleva,Grasso_PPCF}): in the usual, strong guide field limit,  $b$ is completely neglected since is ordered  $b\sim \epsilon^2$ with $\epsilon\equiv |\nabla\psi|/B_z\ll 1$, and to leading order ($\sim\epsilon$) both ${\bm u}_e$ and ${\bm u}_i$ are given by the incompressible ${\bm E}\times{\bm B}$-drift velocity. As consequence, the stream function $\varphi$ corresponds to the normalized electrostatic potential while the $\rho_s^2$ term appears in the electron momentum equation as a result of the diamagnetic corrections to ${\bm u}_e$ in the Lorentz force and the $z$ component of the gyrotropic electron pressure tensor \citep{Schep}. For this reason this term is considered to be an FLR-type contribution. However, the cancellation between the diamagnetic drift contribution to the $z$-component of ${\bm u}_e\cdot\nabla{\bm u}_{e}$ and the $z$-component of the gyrotropic pressure tensor is required in the derivation only if we do not order $\rho_s$ and $d_e$ with respect to $\epsilon$; in that case Eqs.(\ref{eq:OhmRMHD}-\ref{eq:RMHD_U}) contain terms up to the second order in $\epsilon$. If instead we remember that in the slab, strong guide field, RMHD ordering, $\beta_e\sim\epsilon$ and that $\rho_s^2=\beta_ed_i^2/2$, then we may order $\rho_s^2\sim d_e^2\sim\epsilon$. This is sufficient to re-obtain Eqs.(\ref{eq:OhmRMHD}-\ref{eq:RMHD_U}) even by assuming a scalar electron pressure tensor, if  we disregard any contribution of order $\epsilon^4$ or higher, since from ${\bm u}_{e,\perp}\simeq {\bm E}\times{\bm B}/B^2 +{\bm \nabla}P_e\times{\bm B}/(eB^2)$ we would obtain $({\bm u}_e\times{\bm B})\cdot{\bm e}_z=[\varphi-\rho_s^2 U, \psi]$; our equations will now retain terms up to $\epsilon^3$.\\
In  EMHD, instead,  the convection velocity field (i.e. ${\bm u}_\perp^e$) appearing in the second term of  Eq.(\ref{eq:OhmEMHD}) is due to the magnetic field component  $b$, since the current density is carried by electrons only, which  drive the dynamics through ${\bm u}_e\propto{\bm J}\propto{\bm\nabla}\times{\bm B}$ in the incompressible regime that we consider here. As a consequence, $b$ acts as a stream function for the in-plane electron dynamics, and resistivity, when included, enters also in the equivalent of the vorticity equation.  For the same reason, the in-plane components of electron momentum  equation, taken in the polytropic, incompressible limit, completely close the system of EMHD equations: Eq.(\ref{eq:EMHD_U}) is the $z$-component of the rotational of Eq.(\ref{eq:electron}), and the field $W$ is proportional to the $z$-component of the electron generalized vorticity, defined by the curl of the electron fluid canonical momentum ${\bm\nabla}\times({\bm u}_e + e{\bm A}/(m_ec))$.

The EMHD equation for the electron generalized vorticity is mirrored in RMHD by the equation for the fluid vorticity alone (Eq.(\ref{eq:RMHD_U})), of which $\nabla^2\varphi$ represents the $z$-component (see also \citep{Rogers}). This happens because in the Alfv\'enic frequency range the plasma moves at the bulk velocity ${\bm U}\simeq {\bm u}_i+O(m_e/m_i)$): Eq.(\ref{eq:RMHD_U}) is therefore the curl of Eq.(\ref{eq:ion}), under the assumption of incompressibility, which allows expression of the perpendicular fluid velocity in terms of the stream function $\varphi$. If the plasma fluid is assumed to be incompressible but without imposing the strong guide field condition, this function can not be interpeted as the electrostatic potential. With no guide field  however a separate analysis would be required to include $\rho_s^2$-type contributions. The delicate point about the applicability of Eqs.(\ref{eq:OhmRMHD})-(\ref{eq:EMHD_U}) lies indeed in the validity of the incompressibility assumption and in its relationship with the ordering of the parallel fluctuations of the magnetic field, which  weighs the importance of Hall's term in Ohm's law, discussed below.

\subsection{Comparison with the generalized Ohm's law and Hall's term}\label{Ohm_Hall}
Since reconnection models are usually discussed in relation to the non-ideal terms in Ohm's law rather than in the framework of the full two-fluid equations for ions and electrons, it is worth to make here reference also to  the generalized Ohm's law, written with respect to the average plasma velocity ${\bm U}$. The standard text-book form obtained by combining Eqs.(\ref{eq:electron})-(\ref{eq:ion}) (see e.g. \citep{Krall}, p.91) while neglecting  $O(m_e/m_i)$ corrections, reads, after normalizing again lengths to $a$ and times to $\tau_{_A}$,
$${\bm E}+{\bm U}\times {\bm B}=  
 d_i\frac{{\bm J}\times {\bm B}}{n} + S^{-1}{\bm J}
 $$
\begin{equation}\label{eq:Ohm}
+\frac{d_e^2}{n}\left\{\frac{\partial {\bm J}}{\partial t} +{\bm\nabla}\cdot\left(
{\bm U}{\bm J}+ {\bm J}{\bm U}-d_i\frac{{\bm J}{\bm J}}{n} \right)\right\}-\frac{\rho_s^2}{d_i}\frac{{\bm\nabla}\cdot{\bm \Pi}_e}{n}\, .
\end{equation}
 Here $n=n_e=n_i$ is the average plasma density and ${\bm\Pi}_e$ is the electron pressure tensor of Eq.(\ref{eq:electron}), measured in the electron rest frame. The ion pressure tensor contribution is neglected since it is $O(m_e/m_i)$ smaller when the temperatures of the two species are comparable. Note that it has been recently shown by  \emph{Kimura et al. (2014)} \cite{Kimura}  that the (often neglected) term  ${\bm \nabla }\cdot({\bm J}{\bm J}/n)$ is necessary to respect energy conservation of the 1-fluid system in the collisionless limit ($S^{-1}=0$). 

The generalized Ohm's law is essentially the rewriting of the electron momentum equation with respect to ${\bm U}$ and ${\bm J}$,  that replace  ${\bm u}_e$. We then recognize  the essential difference between the dynamics of the bulk plasma and of the magnetic induction, and the role that the Hall-term ${\bm J}\times{\bm B}$ has in this: while the plasma always moves at the fluid velocity of ions, the magnetic induction evolves (with the rotational of Eq.(\ref{eq:Ohm})) as dragged by the fluid velocity of the electrons, ${\bm u}_e=({\bm U}-d_i{\bm J}/n)$. In particular, the term  ${\bm u}_e\times {\bm B}$ describes the convection of magnetic field lines by the electron fluid in the collisionless limit neglecting electron inertia.  As well known \citep{Fruchtman}, the RMHD and EMHD sets of equations for slab reconnection without electron temperature effects may be therefore seen as two extreme limits with respect to the Hall term ($d_i$-term), in Ohm's law: the RMHD regime described by Eqs.(\ref{eq:OhmRMHD}-\ref{eq:RMHD_U}) at $\rho_s=0$  corresponds to neglecting Hall's term entirely, whereas the EMHD framework is  recovered when the fluid dynamics is restricted to electrons only (${\bm U}\simeq{\bm u}_i\simeq 0$), that is at scales $\ell\ll d_i$ and $\Omega_i\lesssim \omega\ll \Omega_e$, so that Eq.(\ref{eq:Ohm}) becomes the only relevant equation for our fluid system. It is however interesting to remark that in the strong guide field ordering, both ions  and electrons in-plane velocities  are equal at the leading order in $\epsilon$ to the ${\bm E}\times{\bm B}$-drift. By direct comparison of the $z$-component of  ${\bm u}_e\times{\bm B}=({\bm U}-d_i{\bm J}/n)\times{\bm B}$ with $[\varphi-\rho_s^2U,\psi]$ (cfr. previous Section), is immediate to recognize that the Hall term survives in the ordering  with $\rho_s^2\sim \epsilon$ through the diamagnetic-drift contribution to ${\bm u}_{e,\perp}$, $\rho_s^2[U,\psi]=d_i({\bm J}\times {\bm B})\cdot{\bm e}_z/n$. This expresses the balance between kinetic and magnetic pressure forces not only at equilibrium but also for the perturbations.

We conclude by recalling that, when Hall's term is retained while still considering the bulk plasma response to field evolution (i.e. ion momentum equation is not neglected, so that ${\bm J}\neq-ne{\bm u}_e$),  an intermediate regime is entered, which is sometimes called  ``Hall-mediated reconnection'' (HMR)  or even ``whistler mediated reconnection\footnote{ Not to be here confused with the EMHD regime of Eqs.(\ref{eq:OhmEMHD})-(\ref{eq:EMHD_U}), though there has been some ambiguous notation for different regimes in the past. Also note that, in some  works, what we here name (resistive) HMR   was even refered to as the ``collisionless reconnection'' regime (see e.g. \citep{Zweibel}), due to the weak dependence from $S$ found in the Hall-dominated reconnection rate (see e.g. \citep{GEM}). }''  \citep{Mandt}. These regimes are not of concern in this paper, since they can not be recovered in the framework of two-field models. The decoupling of ion and electron motions  at the ion inertial scale (i.e. for $\ell\lesssim d_i$) requires more than two scalar fields to be retained to account for two-fluid effects (also notice that Eqs.(\ref{eq:OhmRMHD})-(\ref{eq:EMHD_U}) do not contain $d_i$ as a characteristic scale length). As discussed by \emph{Fruchtman et al. (1993)} \cite{Fruchtman_2}, first, and more recently by \emph{Bian et al. (2007)} \cite{Bian} and  \emph{Hosseinpur et al. (2009)} \cite{Hosseinpour_1}, Hall term effects are retained by relating the magnitude of $b$, as generated by Hall's term in Eq.(\ref{eq:Ohm}), to the  compressible component of ${\bm U}_\perp$, absent in our incompressible model. By Helmholtz decomposition, this should enter through an irrotational contribution, ${\bm U}_\perp={\bm\nabla}\varphi\times {\bm e}_z+{\bm \nabla}\chi$, related to the scalar field $\chi$; in turn, the components $u_z^e$ and $U_z$ should also be retained.  This immediately highlights the most delicate point concerning the ${\bm J}\times{\bm B}$ term in Ohm's law, already pointed out at the end of the previous Section: due to the direct relation between $b$ and $\chi$, the (in)compressibility assumption plays a major role in determining the extent of Hall physics retained in the model. Remarkably, if $\partial_z=0$, the in-plane incompressibility ${\bm\nabla}\cdot{\bm U}_\perp=0$ is admitted both in the ${\bm E}\times{\bm B}$-drift regime of the low-$\beta$ limit, where $b$ is neglected with respect to the strong guide field, and in the high-$\beta$ limit, where the large kinetic (electron) pressure implies the smallness of both ${\bm\nabla}\cdot{\bm U}=0$ and ${\bm\nabla}\cdot{\bm u}^e=0$.

%
%
%
%

\begin{acknowledgments}
The authors are grateful to Francesco Pegoraro for discussions and comments.
 DDS is in debt with Maurizio Ottaviani for many interesting discussions,
 and in particular for having pointed out the possible importance of the ideal tearing during the non-linear stage of primary reconnection instabilities, and with Alessandro Biancalani
for discussions about the explosive reconnection regime and for having kindly provided details about the numerical simulations performed in \citep{Biancalani_EPS,Biancalani}. This research was partially
supported by the joint training PhD program in Astronomy, Astrophysics
and Space Science between the University of Rome ``Tor Vergata'' and ``Sapienza''.
\end{acknowledgments}

%
%
%
%
%
%
%
%
%

\begin{table*}\label{Table_1}
\caption{Characteristic plasma parameters of magnetized plasma environments where MHD reconnection may occur.  Physical quantities are expressed in cgs units and temperatures are expressed in $eV$. For magnetotail reconnection parameters, typical conditions in the plasma sheet during a substorm growth phase have been considered.  For the tokamak devices the value are estimated from design (ITER) or measurements (JET) near to the $q=1$ surface, whose circumference on a poloidal section gives an estimation of the typical, reconnecting current sheet length, $L$.  Source for the parameters, as labelled in the Table's third row, are: \citep{ Shibata2} (I); \citep{Ang, Sergeev, Kivelson} (II); \citep{Porcelli_ITER,JET} (III); \citep{Yamada_MRX} (IV).}
\begin{tabular}{||c||c||c||c|c||c||}\hline
 & \mbox{Low Corona} &     
\mbox{Magnetotail} &
\multicolumn{2}{c||}{\mbox{Tokamak }}  & {\mbox{MRX device}}    \\ 
\hline{~~~~~~}
   &   (\mbox{Sun at} $\sim 1R_\odot$ ) 
    & (\mbox{Central plasma sheet})  & 
\mbox{ITER} & \mbox{JET} &   \\
\hline 
\mbox{Sources:} &  I      & 
II & 
\multicolumn{2}{c||}{III}
 & IV   \\
 \hline
\hline
$L$   & $10^9-10^{10}$   
     & $10^9-10^{10}$  & $900$ & $50$ & $10-20$\\
\hline
$n_e$   & $10^{9}-10^{10}$    & $0.1-1$ & $10^{14}$ & $10^{13}$ & $(2-6)\times 10^{13}$ \\
\hline
$B$   &  $10-100$    & $ 10^{-4}$  & $5.68\times 10^4$ & $3.45\times 10^4$  & $(1-3)\times 10^{2}$ \\
\hline
$T_e$   &  $86$   & $ 10^3-10^4$  & $2\times 10^4$ & $3\times 10^3$  & $5-15$\\
\hline
$\varepsilon_{_S}^*\equiv (S^{-1})^*$   &  $10^{-15}-10^{-12}$  
 &   $10^{-16}-10^{-13}$   & $10^{-11}$ & $10^{-9}$  & $8\times 10^{-4}-3\times 10^{-2}$  \\
\hline
$\varepsilon_d^*\equiv (d_e/L)^2$   & $ 10^{-19}- 10^{-16}$    &
  $10^{-9}-10^{-6}$  & $10^{-9}$ & $10^{-5}$  &  $10^{-5}-10^{4}$\\
\hline
\end{tabular}
\end{table*}

\end{document}